\documentstyle [12pt]{article}

\def\b{\begin{equation}} \def\e{\end{equation}}
\def\bd{\begin{displaystyle}} \def\ed{\end{displaystyle}}
\def\ba{\begin{array}} \def\ea{\end{array}}

\def\bee{\begin{enumerate}}
\def\eee{\end{enumerate}}

\def\bes{\begin{eqnarray*}}
\def\ees{\end{eqnarray*}}
\def\be{\begin{eqnarray}}
\def\ee{\end{eqnarray}}

\begin{document}

\title{Scalar effective action in Krein space quantization}

\author{A. Refaei$^{1,2}$\thanks{e-mail:
abr412@gmail.com}, M.V. Takook$^1$\thanks{e-mail:
takook@razi.ac.ir}}

\maketitle \centerline{\it $^1$Department of Physics, Razi
University, Kermanshah, Iran} \centerline{\it $^2$Islamic Azad
University, Sanandaj branch, Sanandaj, Iran.}

\begin{abstract}

In this paper, the $\lambda\phi^4$ scalar field effective action, in the one-loop approximation, is
calculated by using the Krein space quantization. We show
that the effective action is naturally
finite and the singularity does not appear in the theory. The physical interaction mass, the running coupling
constant and $\beta$-function are then calculated. The effective
potential which is calculated in the Krein space quantization is different from
the usual Hilbert space calculation, however we show that
$\beta$-function is the same in the two different methods.

\end{abstract}
\vspace{0.5cm} {\it PACS numbers}: 04.62.+v, 03.70+k, 11.10.Cd,
98.80.H \vspace{0.5cm}

\section{Introduction}

The quantum gravity is one of the most important problem in theoretical physics.
The linear quantum gravity in the background field method is perturbatively non-renormalizable and also there appear an infrared divergence.
This infrared divergence does not manifest itself in the quadratic part of the
effective action in the one-loop approximation. This means that
the pathological behavior of the graviton propagator may be  gauge
dependent and so should not appear in an effective way as a
physical quantity \cite{anilto}. The infrared
divergence which appears in the linear gravity in de Sitter space
is the same as the minimally coupled scalar field in de Sitter space
\cite{gagarota,derotata}. It is shown that one can not construct a covariant quantization of
the minimally coupled scalar field with only positive norm states
\cite{al}. It has been proved that the use of the two sets of
solutions (positive and negative norms states) is an unavoidable
feature if one wants to preserve causality (locality), covariance
and elimination of the infrared divergence in quantum field theory
for the minimally coupled scalar field in de Sitter space
\cite{gareta,ta3}, {\it i.e.} Krein space quantization.

The singular behavior of Green function at short relative distances
(ultraviolet divergence) or in the large relative distances
(infrared divergence) leads to main divergences in the quantum
field theory. It was conjectured that quantum metric fluctuations
might smear out the singularities of Green functions on the light
cone, but it does not remove other ultraviolet divergences
\cite{for2}. However, it has been shown that quantization in Krein
space removes all ultraviolet divergences of quantum field theory
(QFT) except the light cone singularity \cite{ta4}.  By using the
Krein space quantization and the quantum metric fluctuations
in the linear approximation, we showed that the infinities in the
Green function are disappeared \cite{for2,rota}.

Quantization in Krein space instead of Hilbert
space has some interesting features. For example in this method,
the vacuum energy becomes zero naturally, so the normal ordering
would not be necessary \cite{gareta,ta4}. The auxiliary negative norm states, which are used in the Krein
space quantization, play the regularization  of the theory.

In the present work, using the Krein space method and quantum
metric fluctuation at the linear approximation, we calculate the
one loop effective action for scalar field. It has been shown that
this effective action is naturally regularized. This effective
action differs with what previously reported \cite{itzu}, however
the $\beta$-functions are exactly the same. The effective mass and
coupling constant are also calculated in this method. In this
approximation, we see that this quantization eliminates the
singularity in the theory without changing the $\beta$-functions. In the appendix, details of our
calculation have been presented.

\section{Scalar Green function}

In this section, we review the elementary facts about Krein space
quantization. A classical scalar field $\phi(x)$ satisfies the
following field equation \b
(\Box+m^2)\phi(x)=0=(\eta^{\mu\nu}\partial_\mu
\partial_\nu+m^2)\phi(x),\;\;
\eta^{\mu\nu}=\mbox{diag}(1,-1,-1,-1).\e Inner or {\it
Klein-Gordon} product and related norm are defined by \cite{bida}
\b
(\phi_1,\phi_2)=-i\int_{t=\mbox{const.}}\phi_1(x)\stackrel{\leftrightarrow}
{\partial}_t\phi_2^*(x)d^3x.\e  Two sets of solutions are given
by: \b u_p(k,x)=\frac{e^{i\vec k.\vec x-iwt}}{\sqrt{(2\pi)^32w}}
=\frac{e^{-ik.x}}{\sqrt{(2\pi)^32w}},\;\;u_n(k,x)=\frac{e^{-i\vec
k.\vec x+iwt}}{\sqrt{(2\pi)^32w}}
=\frac{e^{ik.x}}{\sqrt{(2\pi)^32w}},\e where $ w(\vec k)=k^0=(\vec
k .\vec k+m^2)^{\frac{1}{2}} \geq 0$, note that $u_n$ has the
negative norm. In Krein space the quantum field is defined as
follows \cite{ta4} \b \phi(x)=\frac{1}{\sqrt 2}[\phi_p(x)+\phi_n(x)],\e where
$$ \phi_p(x)=\int d^3\vec k [a(\vec k)u_p(k,x)+a^{\dag}(\vec
k)u_p^*(k,x)],$$ $$ \phi_n(x)=\int d^3\vec k [b(\vec
k)u_n(k,x)+b^{\dag}(\vec k)u_n^*(k,x)].$$  $a(\vec k)$ and $b(\vec
k)$ are two independent operators. The time-ordered product
propagator for this field operator is \b iG_T(x,x')=<0\mid
T\phi(x)\phi(x') \mid 0>=\theta (t-t'){\cal W}(x,x')+\theta
(t'-t){\cal W}(x',x).\e In this case we obtain \b
G_T(x,x')=\frac{1}{2}[G_F(x,x')+(G_F(x,x'))^*]=\Re G_F(x,x'),\e
where the Feynman Green function is defined by \cite{bida}$$
G_F(x,x')=\int \frac{d^4 p}{(2\pi)^4}e^{-ip.(x-x') }\tilde
G_F(p)=\int \frac{d^4
p}{(2\pi)^4}\frac{e^{-ip.(x-x')}}{p^2-m^2+i\epsilon}$$ \b
=-\frac{1}{8\pi}\delta
(\sigma_0)+\frac{m^2}{8\pi}\theta(\sigma_0)\frac{J_1
(\sqrt{2m^2\sigma_0})-iN_1 (\sqrt{2m^2\sigma_0})}{\sqrt{2m^2
\sigma_0}}-\frac{im^2}{4\pi^2}\theta(-\sigma_0)\frac{K_1
(\sqrt{-2m^2\sigma_0})}{\sqrt{-2m^2 \sigma_0}},\e in which
$\sigma_0=\frac{1}{2}(x-x')^2 .$ So we have \b G_T(x,x')=\int
\frac{d^4 p}{(2\pi)^4}e^{-ip.(x-x')}{\cal
PP}\frac{1}{p^2-m^2}=-\frac{1}{8\pi}\delta
(\sigma_0)+\frac{m^2}{8\pi}\theta(\sigma_0)\frac{J_1
(\sqrt{2m^2\sigma_0})}{\sqrt{2m^2 \sigma_0}}, \;\;x\neq x',\e
${\cal PP}$ stands for the principal parts. Contribution of the
coincident point singularity $(x=x')$ merely appears in the
imaginary part of $G_F$ (\cite{ta3} and equation (9.52) in \cite{bida})
$$ G_F(x,x)=-\frac{2i}{(4\pi)^2}\frac{m^2}{d-4}+G_F^{\mbox{finit}}(x,x),$$ where $d$ is the space-time dimension and
$G_F^{\mbox{finit}}(x,x)$ becomes finite as $d\longrightarrow 4$.
Note that the singularity of the Eq. $(2.8)$ takes place only on
the cone \emph{i.e.,} $x\neq x', \sigma_0=0$.

It has been shown that the quantum metric fluctuations remove the
singularities of Green's functions on the light cone \cite{for2}.
Therefore, the quantum field theory in Krein space, including the
quantum metric fluctuation $\left(
g_{\mu\nu}=\eta_{\mu\nu}+h_{\mu\nu}\right)$, removes all the
ultraviolet divergencies of the theory \cite{rota,for2}, so one
can write: \b \langle G_T(x - x')\rangle = -\frac{1 }{8\pi}
\sqrt{\frac{\pi}{2\langle\sigma_1^2\rangle}}
exp\left(-\frac{\sigma_0^2}{2\langle\sigma_1^2\rangle}\right)+
 \frac{m^2}{8\pi}\theta(\sigma_0)\frac{J_1(\sqrt {2m^2
 \sigma_0})}{\sqrt {2m^2 \sigma_0}},\e where $2\sigma= g_{\mu\nu}(x^{\mu}-
x'^{\mu})(x^{\nu} - x'^{\nu})$. In the case of $2\sigma_0 =
\eta_{\mu\nu}(x^{\mu}- x'^{\mu})(x^{\nu} - x'^{\nu})=0$, due to
the quantum metric fluctuation ($h_{\mu\nu}$), we have
$\langle\sigma_1^2\rangle\neq 0$ so we get \b \langle
G_T(0)\rangle = -\frac{1 }{8\pi}
\sqrt{\frac{\pi}{2\langle\sigma_1^2\rangle}} +
 \frac{m^2}{8\pi}\frac{1}{2}.\e
It should be noted that $ \langle\sigma_1^2\rangle $ is related to
the density of gravitons \cite{for2}.

By using the Fourier transformation of Dirac delta function,
$$ -\frac{1}{8\pi}\delta(\sigma_0)= \int \frac{d^4
p}{(2\pi)^4}e^{-ip.(x-x')}{\cal PP}\frac{1}{p^2},$$or equivalently
$$\frac{1}{8\pi^2}\frac{1}{\sigma_0}= -\int \frac{d^4
p}{(2\pi)^4}e^{-ip.(x-x')}\pi\delta(p^2),$$ for the second part of
Green function, we obtain \b
\frac{m^2}{8\pi}\theta(\sigma_0)\frac{J_1(\sqrt {2m^2
 \sigma_0})}{\sqrt {2m^2 \sigma_0}}=\int \frac{d^4
p}{(2\pi)^4}e^{-ip.(x-x')}{\cal PP}\frac{m^2}{p^2(p^2-m^2)}.\e
And for the first part we have $$
-\frac{1}{8\pi}\sqrt{\frac{\pi}{2\langle\sigma_1^2\rangle}}exp\left[-\frac{(x-x')^4}
{4\langle\sigma_1^2\rangle}\right]=\int \frac{d^4
p}{(2\pi)^4}e^{-ik.(x-x') }  \tilde{G}_1(p).$$ Therefore, we
obtain \b <\tilde G_T(p)>=\tilde{G}_1(p)+{\cal
PP}\frac{m^2}{p^2(p^2-m^2)}.\e In the previous paper, we proved
that in the one-loop approximation, the Green function in Krein
space quantization which appears in the transition amplitude is
\cite{ta4}: \b <\tilde G_T(p)>\mid_{\mbox{one-loop}}\equiv \tilde
G_T(p)\mid_{\mbox{one-loop}}\equiv {\cal PP}
\frac{m^2}{p^2(p^2-m^2)}. \e That means in the one loop
approximation, the contribution of delta function is negligible.
It is worth to mention that in order to improve the UV behavior in
relativistic higher-derivative correction theories, the propagator
$(2.13)$ has been used by some authors \cite{bach,ho}. It is also appear in supersymmetry (equation (20.76) in \cite{ka}).

\section{Scalar effective potential}

The effective action in the one-loop approximation for $\lambda
\phi^4$ scalar field is defined by \cite{bida,itzu}
$$\Gamma(\phi)=I(\phi)+\frac{i}{2}\hbar\;
Tr\;\ln\left[1+(\Box+m^2)^{-1}V''(\phi)\right]+O(\hbar^2)$$ \b
=I(\phi)+\frac{i}{2}\hbar\; Tr\;\ln\left[1-G_F
V''(\phi)\right]+O(\hbar^2),\e where $G_F$ is the Feynman Green function and
$$ Tr\;\ln\left[1-G_F V''(\phi)\right]=\int d^4x <x\mid \ln\left[1-G_FV''(\phi)\right]\mid x>.$$
By using the Fourier transformation, one obtains $$
Tr\;\ln\left[1-G_FV''(\phi)\right]=\int d^4x\int
\frac{d^4p}{(2\pi)^4}\ln\left[1-V''(\phi)\left(
\frac{1}{p^2-m^2+i\epsilon} \right)\right].$$
The effective potential is
\b  V_{eff}=V_{eff}^{(0)}+\hbar V_{eff}^{(1)}+\hbar^2 V_{eff}^{(2)}+...,
 $$
where $$V_{eff}^{(0)}=\frac{m^2\phi^2}{2}+\frac{\lambda\phi^4}{4!}
,$$ $$ V^{(1)}_{eff}=-\frac{i}{2}\int
\frac{d^4p}{(2\pi)^4}\ln\left[1-\frac{V''(\phi)}{p^2-m^2+i\epsilon}
\right].\e There are two different types of singularity in the
one-loop effective potential \cite{itzu}
$$ V^{(1)}_{eff}=\frac{1}{32\pi^2}\left(-\frac{m^2\lambda
\phi^2}{2}\Gamma(-1)+\frac{1}{2}\left(\frac{\lambda
\phi^2}{2}\right)^2\Gamma(0)\right)$$
\b+\frac{1}{64\pi^2}\left[\left(\frac{\lambda
\phi^2}{2}+m^2\right)^2\ln\left(1+\frac{\lambda \phi^2}{2m^2
}\right)-\frac{\lambda \phi^2}{2}\left(m^2+\frac{3\lambda
\phi^2}{4}\right)\right].\e

By using the Green function $(2.13)$, we have \b
V^{(1)}_{eff}=\frac{-i}{2}\int
\frac{d^4p}{(2\pi)^4}\ln\left[1-V''(\phi){\cal PP}\left(
\frac{m^2}{p^2(p^2-m^2)} \right)\right],\e after  some
calculations (Appendix), we reach the following form for the
effective action in the one loop approximation:
$$ V^{(1)}_{eff}=-\frac{m^2\lambda\phi^2}{(16\pi)^2}\left[\left(1+\frac{\lambda \phi^2}{m^2}\right) \left(-2\ln(1+\frac{\lambda \phi^2}{m^2})+
\ln(1+\frac{\lambda \phi^2}{2m^2})+\ln\frac{\lambda
\phi^2}{2m^2}+2\ln2\right)\right. $$ \b \left.+
\ln(1+\frac{\lambda \phi^2}{2m^2})- \ln\frac{\lambda
\phi^2}{2m^2}\right]-\frac{m^2\lambda\phi^2}{64\pi}
\left[1+\sqrt{1+\frac{2m^2}{\lambda\phi^2}}-2\sqrt{1+\frac{m^2}{\lambda\phi^2}}\right].\e
So, it is clear that in this method, the effective potential does not
contain any divergence. In other words, in the Krein space method, the effective action is automatically regularized. \\ Now let's consider the effective mass and coupling constant. They are
defined by
$$ m_{eff}^2(\mu)=\left.\frac{d^2V_{eff}}{d\phi^2}\right| _{\phi=\mu},$$
$$ \lambda_{eff}(\mu)=\left. \frac{d^4V_{eff}}{d\phi^4}\right| _{\phi=\mu}.$$
For our effective potential in $\mu\rightarrow0$, we obtain
$$m_{eff}^2=m^2\left(1-\frac{\lambda}{32\pi}-\frac{\lambda}{64\pi^2}\ln2 \right)+O(\lambda^2),$$
where $m_{eff}$ and $m$ are measurable quantities, they can be interpreted as the physical particle interaction mass and the physical free particle mass.

$\lambda_{eff}$ is a coupling constant in
the presence of interaction which is finite. It is a function of
a constant $\mu$ which is the energy scale of the
interaction,
$$\lambda_{eff}\equiv\lambda_{\mu}=\lambda
-\frac{\lambda^2}{(8\pi)^2}\left[6\ln\frac{\mu^2}{m^2}+19+12\ln2\right]+O(\lambda^3).$$
By defining  $\mu=e^{-t}$ and the running coupling constant
$\bar\lambda(t,\lambda)$, the Beta function is \b
\beta=\frac{d\bar\lambda(t,\lambda)}{dt}=\frac{3\lambda^2}{16\pi^2}.\e
Our potential is different from previous methods but the
interesting point is that in the one-loop approximation the
$\beta$ function does not change.

\section{Conclusion and outlook}

We recall that the negative frequency solutions of the field
equation will be needed for quantizing in a correct way like the minimally
coupled scalar field in de Sitter space. Contrary to the Minkowski
space, the elimination of de Sitter negative norm in the minimally
coupled states breaks the de Sitter invariance. Then, for restoring
the de Sitter invariance, one needs to take into account the
negative norm states {\it i.e.} the Krein space  quantization. It
provides a natural tool for eliminating the singularity in the QFT
\cite{gareta}.

A theory becomes sensible when it can explain the experimental data for physical
quantities and moreover its capacity for predicting the amount of new quantities which have not
been measured yet. Every new observed quantity that is in agreement with what the theory predicts is a
 support for the theory.

In Quantum Field Theory, calculation of quantum effects of the physical quantities
is made by the means of the expectation values which are related
to the Green's function. Since the divergences of Green's
functions usually appear in one of the following formats:
$$\lim_{\sigma\rightarrow 0}\;\;\;\; \frac{1}{\sigma},\;\; \ln\sigma,\;\; \delta(\sigma),$$ so the
divergence is automatically involved the calculations. It has
been observed that in the Krein space quantization, by considering
metric fluctuations, the divergence of the Green's function is
removed. Therefore this method of quantization can be regarded as
a new method for regularization, (Krein regularization).

After this regularization, one can proceed renormalization according to
the previous method. This new kind of regularization may be
utilized in the calculation of the Lamb-Shift and
Magnetic-Anomaly. If we consider the renormalization point at $
p^2=m^2$, we will exactly obtain the previous reported results for the Lamb-Shift and
Magnetic-Anomaly in the one-loop approximation \cite{zafota}.

In this paper, Krein space quantization has been used to calculate
the effective action for $\lambda \phi^4$ theory in Minkowski
space-time in the one-loop approximation. It is found that in this
approximation the theory is free of any divergence since the Green
function is free of any divergence in the ultraviolet and infrared
limit. Our potential is different from previous methods but in the
one-loop approximation, the $\beta$ function does not change. In
this approximation and for scalar field, we see that this
quantization eliminates the singularity in the theory without
changing the physical content of the theory. As a future work, one
may use this method for considering the quantum gravity in
the background field method. This method may solved
the non-renormalizability of quantum gravity in the background
field method.

\vskip 0.5 cm

\noindent {\bf{Acknowledgments}}: The author would like to thank
M. R. Tanhayi.

\begin{appendix}
\setcounter{equation}{0}
\section{Appendix}

In this appendix, we explicitly calculate the integral $(3.17)$. We have
$$ V_{eff}^{(1)}=\frac{-i}{2}\int
\frac{d^4p}{(2\pi)^4}\ln\left[1-V''(\phi){\cal PP}\left(
\frac{m^2}{p^2(p^2-m^2)} \right)\right]=$$
$$ \frac{-i}{2}\int
\frac{d^4p}{(2\pi)^4}\ln\left[1-\frac{m^2V''(\phi)}{2}\left(
\frac{1}{p^2(p^2-m^2)+i\epsilon}+ \frac{1}{p^2(p^2-m^2)-i\epsilon}
\right)\right]=$$
   \b \frac{-i}{2}\int \frac{d^4p}
{(2\pi)^4}\ln
\left[\left(1-\frac{\frac{m^2\lambda\phi^2}{4}}{p^2(p^2-m^2)-i\epsilon
}\right)\left(1-\frac{\frac{m^2\lambda\phi^2}{4}}{p^2(p^2-m^2)+i\epsilon
}\right)-\left(\frac{\frac{m^2\lambda\phi^2}{4}}{p^2(p^2-m^2)}\right)^2
\right],\e where $\epsilon^2$ has been vanished. One can write $$\frac{-i}{2}\int
\frac{d^4p} {(2\pi)^4}
 \ln \left[ \left(1-\frac{\frac{m^2\lambda\phi^2}{4}}{p^2(p^2-m^2)-i\epsilon
}\right)
\left(1-\frac{\frac{m^2\lambda\phi^2}{4}}{p^2(p^2-m^2)+i\epsilon
}\right)\right.$$ \b \left.
 \left(1-\frac{(\frac{\frac{m^2\lambda\phi^2}{4}}{p^2(p^2-m^2)})^2}{(1-\frac{\frac{m^2\lambda\phi^2}{4}}{p^2(p^2-m^2)
 -i\epsilon
})(1-\frac{\frac{m^2\lambda\phi^2}{4}}{p^2(p^2-m^2)+i\epsilon
})}\right)\right].\e So we have
     \b \frac{-i}{2}\int
\frac{d^4p}{(2\pi)^4}
\left[\ln\left(1-\frac{\frac{m^2\lambda\phi^2}{4}}{p^2(p^2-m^2)-i\epsilon
}\right)+ \frac{-i}{2}\ln
\left(1-\frac{\frac{m^2\lambda\phi^2}{4}}{p^2(p^2-m^2)+i\epsilon
}\right)\right]+
$$ $$\frac{-i}{2}\int
\frac{d^4p}{(2\pi)^4}\ln\left[1-\left(\frac{(\frac{\frac{m^2\lambda\phi^2}{4}}{p^2(p^2-m^2)})^2}
{(1-\frac{\frac{m^2\lambda\phi^2}{4}}{p^2(p^2-m^2)-i\epsilon
})(1-\frac{\frac{m^2\lambda\phi^2}{4}}{p^2(p^2-m^2)+i\epsilon
})}\right)\right] .\e
After some calculations we obtain
 \b \frac{-i}{2}\int
\frac{d^4p}{(2\pi)^4}
\ln\left(1-\frac{\frac{m^2\lambda\phi^2}{4}}{p^2(p^2-m^2)-i\epsilon
}\right)+ \frac{-i}{2}\int \frac{d^4k}{(2\pi)^4}\ln
\left(1-\frac{\frac{m^2\lambda\phi^2}{4}}{p^2(p^2-m^2)+i\epsilon
}\right)+$$ $$
\frac{-i}{2}\int\frac{d^4p}{(2\pi)^4}\ln\left[1-\left(\frac{\frac{m^2\lambda\phi^2}{4}}{p^2(p^2-m^2)
-\frac{m^2\lambda\phi^2}{4}}\right)^2\right].
\e

By using Wick rotation $( p_0\rightarrow ik_0)$ , the
corresponding Euclidean four-momentum $k$, we find that the two
first integrals are vanished and the third splits into two
integrals \b\frac{1}{2} \int \frac{d^4k}{(2\pi)^4}
\ln(1-\frac{\frac{m^2\lambda\phi^2}{4}}{k^2(k^2+m^2)-\frac{m^2\lambda\phi^2}{4}})
+ \frac{1}{2}\int \frac{d^4k}{(2\pi)^4}
\ln(1+\frac{\frac{m^2\lambda\phi^2}{4}}{k^2(k^2+m^2)-\frac{m^2\lambda\phi^2}{4}}).\e
Now we want to solve these integrals. The first term is: \b \frac{1}{2}\int
\frac{d^4k}{(2\pi)^4}
\ln(1-\frac{\frac{m^2\lambda\phi^2}{4}}{k^2(k^2+m^2)-\frac{m^2\lambda\phi^2}{4}})=
\frac{1}{2}\int \frac{d^4k}{(2\pi)^4}
\ln(1-\frac{\frac{m^2\lambda\phi^2}{4}}{(k^2+\frac{m^2}{2})^2-\frac{m^4}{4}-\frac{m^2\lambda\phi^2}{4}})=
$$ $$ -\frac{1}{2}\int \frac{d^4k}{(2\pi)^4}
\int_0^{\frac{m^2\lambda\phi^2}{4}}\frac{du}{(k^2+\frac{m^2}{2})^2-\frac{m^4}{4}-\frac{m^2\lambda\phi^2}{4}-u}=
$$ $$ -\frac{1}{2}\int \frac{d^4k}{(2\pi)^4}
\int_0^{\frac{m^2\lambda\phi^2}{4}}du\int_0^\infty dt\exp \left[
{-t(q^2-\frac{m^4}{4}-\frac{m^2\lambda\phi^2}{4}-u})\right],\e
where $ q=k^2+\frac{m^2}{2}$ and it is necessary  $ q^2>
\frac{m^4}{4}+\frac{m^2\lambda\phi^2}{2}.$  So, we see that there
is a cutoff momentum, after doing some straightforward
calculations one obtains $k_c=\sqrt{\frac{\lambda\phi^2}{2}}$ or $
k_c<k<\infty $. We obtain
$$-\frac{1}{32\pi^2}\int_0^\infty dt \left( \int _{q_c}^\infty dq(q-\frac{m^2}{2})e^{-tq^2} \int_0^{\frac{m^2\lambda\phi^2}{4}}due^{tu}\right)\exp \left[ {t(\frac{m^4}{4}+\frac{m^2\lambda\phi^2}{4}})\right]=$$
$$-\frac{1}{32\pi^2}\int_0^\infty dt \left( \frac{1}{2t}e^{-tq_c^2}-\frac{m^2}{4}\sqrt{\frac{\pi}{t}}e^{-tq_c^2}\right)
\left(\frac{1}{t}(e^{\frac{1}{4}tm^2\lambda\phi^2}-1)\right)\exp\left[
t(\frac{m^4}{4}+ \frac{m^2\lambda\phi^2}{4})\right]\equiv A$$
The second term in eq. (A.5) is
 \b \frac{1}{2}\int \frac{d^4k}{(2\pi)^4}
\ln(1+\frac{\frac{m^2\lambda\phi^2}{4}}{k^2(k^2+m^2)-\frac{m^2\lambda\phi^2}{4}})=
\frac{1}{2}\int \frac{d^4k}{(2\pi)^4}
\ln(1+\frac{\frac{m^2\lambda\phi^2}{4}}{(k^2+\frac{m^2}{2})^2-\frac{m^4}{4}-\frac{m^2\lambda\phi^2}{4}})=
$$ $$ \frac{1}{2}\int \frac{d^4k}{(2\pi)^4}
\int_0^{\frac{m^2\lambda\phi^2}{4}}\frac{du}{(k^2+\frac{m^2}{2})^2-\frac{m^4}{4}-\frac{m^2\lambda\phi^2}{4}+u}=
$$ $$ \frac{1}{2}\int \frac{d^4k}{(2\pi)^4}
\int_0^{\frac{m^2\lambda\phi^2}{4}}du\int_0^\infty dt\exp \left[
{-t(q^2-\frac{m^4}{4}-\frac{m^2\lambda\phi^2}{4}+u})\right]\e
$$\frac{1}{32\pi^2}\int_0^\infty dt \left( \int _{q_c}^\infty
dq(q-\frac{m^2}{2})e^{-tq^2}
\int_0^{\frac{m^2\lambda\phi^2}{4}}due^{-tu}\right)\exp \left[
{t(\frac{m^4}{4}+\frac{m^2\lambda\phi^2}{4}})\right]=$$
$$\frac{1}{32\pi^2}\int_0^\infty dt \left(
\frac{1}{2t}e^{-tq_c^2}-\frac{m^2}{4}\sqrt{\frac{\pi}{t}}e^{-tq_c^2}\right)
\left(-\frac{1}{t}(e^{-\frac{1}{4}tm^2\lambda\phi^2}-1)\right)\exp
\left[ {t(\frac{m^4}{4}+\frac{m^2\lambda \phi^2}{4}})\right]\equiv
B $$ By summing A and B we get \b
A+B=-\frac{1}{32\pi^2}\sum_{n=1}\int_0^\infty
\frac{dt}{4^{2n}{(2n)}!}  t^{2n-2}(m^2\lambda\phi^2)^{2n}\exp
\left[ {-t\frac{\lambda\phi^2}{4}(m^2+\lambda \phi^2})\right]+$$
$$\frac{m^2}{64\pi^\frac{3}{2}}\sum_{n=1}\int_0^\infty \frac{dt}{4^{2n}{(2n)}!}t^{2n-\frac{3}{2}}
(m^2\lambda\phi^2)^{2n}\exp \left[
{-t\frac{\lambda\phi^2}{4}(m^2+\lambda \phi^2})\right].\e
Then for the one loop case we have
\b
V_{eff}^{(1)}=-\frac{1}{32\pi^2}\sum_{n=1}\int_0^\infty
\frac{dt}{4^{2n}{(2n)}!}  t^{2n-2}(m^2\lambda\phi^2)^{2n}\exp
\left[ {-t\frac{\lambda\phi^2}{4}(m^2+\lambda \phi^2})\right]+$$
$$\frac{m^2}{64\pi^\frac{3}{2}}\sum_{n=1}\int_0^\infty \frac{dt}{4^{2n}{(2n)}!}t^{2n-\frac{3}{2}}
(m^2\lambda\phi^2)^{2n}\exp \left[
{-t\frac{\lambda\phi^2}{4}(m^2+\lambda \phi^2})\right],\e or
\b
V_{eff}^{(1)}=-\frac{1}{(16\pi)^2}\sum_{n=1}
\frac{m^2\lambda\phi^2}{n(2n-1)}\frac{1}{(1+\frac{\lambda
\phi^2}{m^2})^{2n-1}}+ $$
$$\frac{m^2}{64\pi^\frac{3}{2}}\sum_{n=1}\sqrt{m^2\lambda\phi^2} \frac{(2n-\frac{3}{2})!}{(2n)!}\frac{1}{(1+\frac{\lambda \phi^2}{m^2})^{2n-\frac{1}{2}}}.\e
And finally we have
\b
V_{eff}^{(1)}=-\frac{m^2\lambda\phi^2}{(16\pi)^2}\left[\left(1+\frac{\lambda
\phi^2}{m^2}\right)\ln \left(1-\frac{1}{(1+\frac{\lambda
\phi^2}{m^2})^2}\right)+\ln\left(\frac{2+\frac{\lambda
\phi^2}{m^2}}{\frac{\lambda \phi^2}{m^2}}\right)\right]-$$
$$\frac{m^2\lambda\phi^2}{64\pi}\left[1+\sqrt{1+\frac{2m^2}{\lambda\phi^2}}-2\sqrt{1+\frac{m^2}{\lambda\phi^2}}\right],\e
or $$ V_{eff}^{(1)}=-
\frac{m^2\lambda\phi^2}{(16\pi)^2}\left[\left(1+\frac{\lambda
\phi^2}{m^2}\right) \left(-2\ln(1+\frac{\lambda
\phi^2}{m^2})+\ln(1+\frac{\lambda \phi^2}{2m^2})+\ln\frac{\lambda
\phi^2}{2m^2}+2\ln2\right)\right.$$ \b \left. + \ln(1+\frac{\lambda
\phi^2}{2m^2})-\ln\frac{\lambda \phi^2}{2m^2}\right]
-\frac{m^2\lambda\phi^2}{64\pi}\left[1+\sqrt{1+\frac{2m^2}{\lambda\phi^2}}-2\sqrt{1+\frac{m^2}{\lambda\phi^2}}\right].\e

\end{appendix}


\begin{thebibliography}{a}


\bibitem{anilto} I. Antoniadis, J. Iliopoulos, T.N. Tomaras, Nuclear
Phys. B, 462(1996)437.
\bibitem{gagarota} T. Garidi et al, J. Math. Phys., 49(2008)032501;
T. Garidi et al, J. Math. Phys., 44(2003)3838; S. Behroozi et al, Phys. Rev. D, 74(2006)124014.
\bibitem{derotata} M. Dehghani et al, Phys. Rev. D, 77(2008)064028; M.V. Takook et al,
J. Math Phys., 51(2010)032503.
\bibitem{al} B. Allen, Phys. Rev. D, 32(1985)3136.
\bibitem{gareta} J.P. Gazeau, J. Renaud, M.V. Takook, Class. Quan.
Grav., 17(2000)1415, gr-qc/9904023.
\bibitem{ta3} M.V. Takook, Mod. Phys. Lett. A, 16(2001)1691,
gr-qc/0005020.
\bibitem{ta4} M.V. Takook, Int. J. Mod. Phys. E, 11(2002)509,
gr-qc/0006019.
\bibitem{for2} H.L. Ford, Quantum Field Theory in Curved Spacetime,
gr-qc/9707062.
\bibitem{rota} S. Rouhani, M.V. Takook, Int. J. Theor. Phys., 48(2009)2740–2747.
\bibitem{itzu} C. Itzykson, J.B. Zuber, McGraw-Hill, Inc. (1988) {\it Quantum Field
Theory}.
\bibitem{bida} N.D. Birrell, P.C.W. Davies, Cambridge University Press, (1982)
{\it QUANTUM FIELD IN CURVED SPACE}.
\bibitem{bach}  N.H. Barth, S.M. Christensen,
 Phys. Rev. D, 28(1983)1876.
\bibitem{ho} P. Horava, Phys. Rev. D, 79(2009)084008, arXiv:0901.3775.
\bibitem{ka} M. Kaku, Oxfor University Press, (1993) {\it Quantum Field
Theory}.
\bibitem{zafota} A. Zarei, B. Forghan, M.V. Takook, {\it QED in Krein Regularization}, In preparation (2010).

\end{thebibliography}
\end{document}